\documentclass[epj]{webofc}
\usepackage[utf8]{inputenc}
\usepackage[varg]{txfonts}   % Web of Conferences font
\usepackage{booktabs}
\usepackage{xcolor}
\definecolor{darkred}{rgb}{0.4,0.0,0.0}
\definecolor{darkgreen}{rgb}{0.0,0.4,0.0}
\definecolor{darkblue}{rgb}{0.0,0.0,0.4}
\usepackage[bookmarks,linktocpage,colorlinks,
    linkcolor = darkred,
    urlcolor  = darkblue,
    citecolor = darkgreen]{hyperref}
\usepackage{subfigure}
\wocname{EPJ Web of Conferences}
\woctitle{Lattice2017}
\newcommand{\eq}[1]{eq.~(\ref{#1})}

\newcommand{\fig}[1]{Fig.~\ref{#1}}
\newcommand{\Fig}[1]{Figure~\ref{#1}}

\newcommand{\tab}[1]{Table~\ref{#1}}

\newcommand{\ALPHALogo}{\includegraphics[height=32pt]{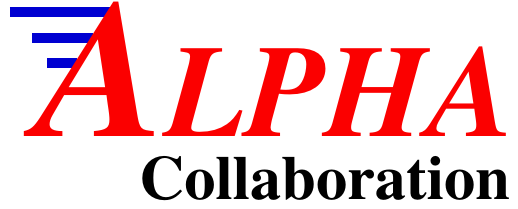}}
\newcommand{\Mc}{M_{\rm c}}
\newcommand{\Nf}{N_{\mathrm{f}}}
\newcommand{\lag}[1]{{\mathcal{L}}_{\rm {#1}}}
\newcommand{\Lsix}{\lag{6}}
\newcommand{\LamYM}{\Lambda_{\rm YM}}
\newcommand{\msbar}{{\rm \overline{MS\kern-0.05em}\kern0.05em}}
\newcommand{\mhad}{m^\mathrm{had}}
\newcommand{\mhadYM}{\mhad_{\rm YM}}
\newcommand{\rmO}{{\rm O}}
\def\fm{{\rm fm}}
\newcommand{\ev}[1]{\left\langle #1 \right\rangle}
\newcommand{\rmd}{{\rm d}}
\newcommand{\rmA}{{\rm A}}
\begin{document}
%%%%%%%%%%%%%%%%%%%%%%%%%%%%%%%%%%%%%%%%%%%%%%%%%%%%%%%%%%%%%%%%%%%%%%%%%%%%%
%
\selectlanguage{english}
%----------------------------------------------------------------------------
\title{%
Decoupling of charm beyond leading order
}
%----------------------------------------------------------------------------
\author{%
\firstname{\ALPHALogo \hfill\\}
\firstname{Francesco} \lastname{Knechtli}\inst{1}\fnsep\thanks{Speaker, \email{knechtli@physik.uni-wuppertal.de}} \and
\firstname{Tomasz} \lastname{Korzec}\inst{1} \and
\firstname{Bj{\"o}rn}  \lastname{Leder}\inst{2} \and
\firstname{Graham}  \lastname{Moir}\inst{3}
% etc.
}
%----------------------------------------------------------------------------
\institute{%
Fakult\"at f\"ur Mathematik und Naturwissenschaften, Bergische Universit\"at Wuppertal, Gau{\ss}str. 20, 42119 Wuppertal, Germany
\and
Institut f\"ur Physik, Humboldt Universit\"at, Newtonstr. 15, 12489 Berlin, Germany
\and
DAMTP, Centre for Mathematical Sciences, Wilberforce Road, CB3 0WA Cambridge, UK
}
%----------------------------------------------------------------------------
\abstract{%
We study the effective theory of decoupling of a charm quark at low energies.
We do this by simulating a model, QCD with two mass-degenerate charm quarks.
At leading order the effective theory is a pure gauge theory. 
By computing ratios of hadronic scales we
have direct access to the power corrections in the effective theory. We
show that these corrections follow the expected leading behavior, which is
quadratic in the inverse charm quark mass.
}
%----------------------------------------------------------------------------
\maketitle
%----------------------------------------------------------------------------
\section{Introduction}\label{s:intro}

Several lattice groups are performing simulations of QCD$_3$
with up, down and strange quarks. They are cheaper and simpler than
simulations of QCD$_4$ which includes a dynamical charm quark.
The motivation to neglect the charm quark at low energies
$E\ll M_{\rm charm}\equiv\Mc$ is decoupling.
QCD$_4$ can be described by an effective theory which at leading order is 
QCD$_3$ without the charm quark.
Neglecting the light quark masses,
the effects of decoupling of the charm at low energies are
incorporated in the matching of the gauge couplings and in the
power corrections stemming from higher order terms in the effective theory.
Matching of the gauge couplings of QCD$_4$ and QCD$_3$ can be performed in 
perturbation theory. This is used in the determination of $\alpha_s$ from
QCD$_3$ simulations by the ALPHA collaboration~\cite{tomaszlat17,Bruno:2017gxd}.
Non-perturbative tests of decoupling are important. On one hand to confirm
the applicability of perturbation theory, on the other hand
to know the size of the power corrections and
whether they can be neglected. We address the latter issue here.

In order to avoid a multi-scale problem and control the continuum limit
we study a model, QCD$_2$ with $\Nf=2$ degenerate quarks of mass $M$.
In this case the effective theory for $E\ll M$ is a Yang--Mills theory 
($\Nf=0$, $M=\infty$) at leading order.
Beyond leading order there are power corrections starting at $M^{-2}$.
In a previous work~\cite{Bruno:2014ufa} simulation of masses up to 
$M\approx \Mc/2$ allowed to estimate that the size of the power corrections
is at the permille level for $M=\Mc$. These estimates were based on
interpolations with the pure gauge theory at $M=\infty$.
However a behaviour of the power corrections $\propto M^{-2}$ was not seen.
In this contribution, which is based on~\cite{Knechtli:2017xgy}, 
we extend the mass range of the simulations to $M\gtrsim \Mc$ thus allowing
a direct determination of the power corrections for a charm quark and a
more stringent test of the behavior expected from the effective theory.

%----------------------------------------------------------------------------
\section{Effective theory of decoupling}\label{s:dec}

In this work we consider only virtual effects of a heavy quark with mass $M$.
We exclude states with explicit heavy quarks from the discussion.
The decoupling of the heavy quark at low energies can be described
in terms of an effective Lagrangian~\cite{Weinberg:1980wa}. 
In the case of decoupling of $\Nf=2$ mass-degenerate heavy quarks it reads
    \begin{eqnarray}
\lag{dec} &=& \lag{YM}
+ \frac{1}{M^{2}} \Lsix +\rmO\left(\frac{\Lambda^{4}}{M^{4}}\right) \,,
\label{e:Ldec}
         \\
         \Lsix &=&\omega_1\,\mathrm{tr}\{D_\mu F_{\nu\rho} D_\mu F_{\nu\rho}\}
+\omega_2\,\mathrm{tr}\{D_\mu F_{\mu\rho} D_\nu F_{\nu\rho}\} \,.
\label{e:Lsix}
    \end{eqnarray}
In \eq{e:Ldec} $\lag{YM}$ is the Yang--Mills Lagrangian which describes the 
leading order in an expansion in inverse powers of $M$.
Due to gauge invariance there are no dimension 5 operators. Therefore
the first correction starts at $M^{-2}$ with the dimension 6 Lagrangian
$\Lsix$ defined in \eq{e:Lsix} which contains two independent 
terms~\cite{Cho:1994yu,Manohar:1997qy}.

The Yang--Mills Lagrangian $\lag{YM}$ has one free parameter, the gauge 
coupling.
Matching the $\Nf=2$ and Yang--Mills theories means specifying a value of 
the Yang--Mills coupling at some scale or equivalently its $\Lambda$ parameter.
Matching can be described by the relation
\begin{equation}\label{e:P}
\LamYM(M,\Lambda) = P(M/\Lambda)\,\Lambda
\end{equation}
which expresses the effective Lambda parameter $\LamYM$ of the Yang--Mills
theory after matching as a function of the heavy quark mass and the Lambda 
parameter of the $\Nf=2$ theory $\Lambda\equiv \Lambda^{(\Nf=2)}$.\footnote{
We use the $\msbar$ scheme for the $\Lambda$ parameters. The heavy quark
mass $M$ is taken to be the renormalization group invariant quark mass
which is the same in all mass-independent renormalization schemes.}
For dimensional reasons this function is a dimensionless factor $P(M/\Lambda)$
times $\Lambda$.

Consider a low energy hadronic observable $\mhad$. 
It can be a hadronic scale such as $1/\sqrt{t_0}$~\cite{Luscher:2010iy} 
or $1/r_0$~\cite{Sommer:1993ce}.
After matching it takes the same value in both theories up to power 
corrections:
\begin{equation}\label{e:matching}
\mhad(M) = \mhadYM + \rmO(\Lambda^{2}/M^{2})
\end{equation}
Note that the value $\mhadYM$ of the observable in the Yang--Mills theory 
depends on $M$ through the matching \eq{e:matching}. 
This mass dependence is described by the factor $P$ in \eq{e:P} since 
$c=\mhadYM/\LamYM$ is a pure number.
Consider now two hadronic scales,
$m^{\mathrm{had,1}}(M)$ and $m^{\mathrm{had,2}}(M)$, whose values in the
Yang--Mills theory are denoted by $m_{\rm YM}^{\mathrm{had,1}}$ and
$m_{\rm YM}^{\mathrm{had,2}}$ respectively. A consequence of the matching 
relation \eq{e:matching} for the ratio of two hadronic scales is
\begin{eqnarray}\label{e:ratio}
R(M) & = & \frac{m^{\mathrm{had,1}}(M)}{m^{\mathrm{had,2}}(M)}
     = \frac{m_{\rm YM}^{\mathrm{had,1}}}{m_{\rm YM}^{\mathrm{had,2}}} + 
       \rmO(\Lambda^{2}/M^{2})
\end{eqnarray}
Note that the ratio of scales in the Yang--Mills theory
\begin{eqnarray}
\frac{m_{\rm YM}^{\mathrm{had,1}}}{m_{\rm YM}^{\mathrm{had,2}}} & = &
\frac{m_{\rm YM}^{\mathrm{had,1}}/\LamYM}{m_{\rm YM}^{\mathrm{had,2}}/\LamYM} =
\frac{c_1}{c_2}
\end{eqnarray}
is given by the ratio of two pure numbers and is independent of the
Lambda parameter $\LamYM$ (or of the gauge coupling). The matching of 
the couplings is therefore irrelevant for the ratios in \eq{e:ratio}. 
An immediate consequence of \eq{e:ratio} is
\begin{equation}\label{e:ratio_EFT}
R(M) = R(\infty) + k \,\Lambda^{2}/M^{2} + \rmO(\Lambda^{4}/M^{4}) \,,
\end{equation}
where $k$ is a number which depends on the ratio $R$.

%----------------------------------------------------------------------------
\section{Model calculations with two heavy quarks}\label{s:sim}

The data of Ref.~\cite{Bruno:2014ufa} were generated from
simulations of $\Nf=2$ O($a$) improved Wilson quarks with plaquette gauge 
action. In this work
we use $\Nf=2$ twisted mass~\cite{Frezzotti:2000nk} Wilson quarks at 
maximal twist with clover term~\cite{Jansen:1998mx} and plaquette gauge action.
We simulated larger masses corresponding to $\Mc/\Lambda=4.8700$ (charm) and 
$M/\Lambda=5.7781$. We also simulated the pure gauge theory 
($\Nf=0$, $M=\infty$). 
The parameters of the new simulations are summarized in \tab{t:ens}.
We use open boundary conditions and the publicly available
{\tt openQCD} simulation package~\cite{algo:openQCD}.\footnote{
{\tt http://luscher.web.cern.ch/luscher/openQCD/}}
More details on the simulations can be found in Ref.~\cite{Knechtli:2017xgy}.
\begin{table}[tph]
\small
 \centering
 \caption{List of ensembles generated with a doublet of twisted mass Wilson 
quarks at maximal twist for masses $M=\Mc$ ($\Mc/\Lambda=4.8700$) and 
$M=1.2\Mc$ ($M/\Lambda=5.7781$). Also listed are the pure gauge ensembles
($\Nf=0$, $M=\infty$). Open (``o'') boundary conditions (BC) are used.}
 \label{t:ens}
\begin{tabular}{cccccccccc}
\toprule 
$\beta$ & $a$ [$\fm$] & A & BC & $T\times L^3$   & $M/\Lambda_\msbar$ & $t_0/a^2$ & kMDU & $\tau_{\rm exp}$ [kMDU] \\
\midrule
$5.6$ & $\approx0.042$ & tm  & o & $192\times48^3$  & 4.8700   & 6.609(15) & 2.0 & 0.08 \\
      &                &     &   & $192\times48^3$  & 5.7781   & 6.181(11) & 2.1 & 0.08 \\
\midrule
$5.7$ & $\approx0.036$ & tm  & o & $120\times 32^3$ & 4.8703   & 9.104(36) & 17.2& 0.14 \\
      &                &     &   & $192\times 48^3$ & 5.7781   & 8.565(31) & 2.7 & 0.12\\
\midrule
$5.88$ & $\approx0.028$ & tm & o & $192\times48^3$  & 4.8700   & 14.622(62) & 23.1 & 0.24\\
       &                &    &   & $120\times 32^3$ & 5.7781   & 14.916(93) & 59.9 & 0.23\\
\midrule
$6.0$  & $\approx0.023$ & tm & o & $192\times48^3$  & 4.8700   & 22.39(12) & 22.4 & 0.36 \\
\midrule
$6.100$& $0.0778$ & -- & o & $120\times 32^3$ & $\infty$       & 4.4329(32) & 64.0 & 0.05\\
$6.340$& $0.0545$ & -- & o & $120\times 32^3$ & $\infty$       & 9.034(29) & 20.1 & 0.13\\
$6.340^*$&$0.0545$& -- & o & $120\times 24^3$ & $\infty$       & 9.002(31) & 60.9 & 0.13\\ 
$6.672$& $0.0350$ & -- & o & $192\times48^3$  & $\infty$       & 21.924(81) & 73.9 & 0.35\\
$6.900$& $0.0261$ & -- & o & $192\times64^3$  & $\infty$       & 39.41(15) & 160.2& 0.65\\
\bottomrule
\end{tabular}
\end{table}

The lattice spacing for the $\Nf=2$ theory is determined from the scale 
$L_1/a$~\cite{Blossier:2012qu,Fritzsch:2012wq}. For the $\Nf=0$ theory
we use the scale $r_0/a$. Our spatial box sizes $L$ are such that
$L m_\mathrm{PS}\gg 4$ and $L/\sqrt{t_0}\ge8$. In the simulation marked
by $^*$ we explicitley checked that finite volume effects are negligible.
\begin{figure}[t]
  \centering
  \includegraphics[width=10cm,clip]{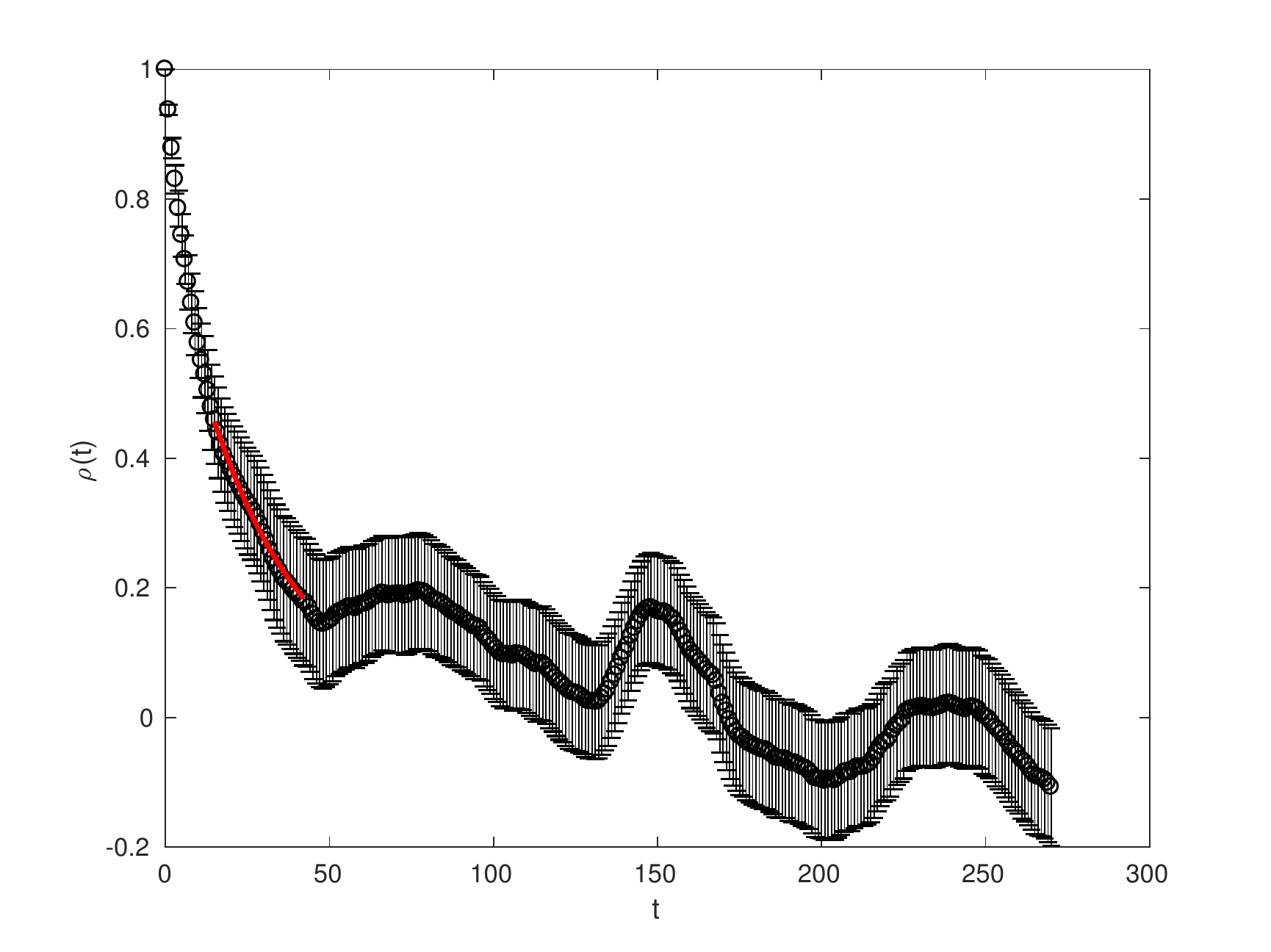}
  \caption{Autocorrelation function of $t_0$ for $\Nf=2$, $\beta=6.0$, $M=\Mc$.
The units on the $x$-axis correspond to 16 Molecular Dynamics Units (MDU). 
The red line is an exponential fit to estimate the exponential autocorrelation 
time.}
  \label{f:tau}
\end{figure}

\Fig{f:tau} shows the autocorrelation function of $t_0$ (in units of 16 MDU) 
for the simulation $\Nf=2$, $\beta=6.0$, $M=\Mc$. The autocorrelation
function for a derived quantity like $t_0$ is defined as in Eq.~(33) of 
Ref.~\cite{Wolff:2003sm}. A fit of the form~\cite{Schaefer:2010hu} 
$A \exp(-t/\tau_{\rm exp})$ to the tail between 
$t=15$ und $t=42$ (represented by the red line in \fig{f:tau}) 
gives an estimate of the
exponential autocorrelation time $\tau_{\rm exp}= 477(101)$ MDU.
Considering all our ensembles we find a behaviour which can be parametrized by
$\tau_{\rm exp} = -32(23) + 17.4(2.8)\,t_0/a^2$.
The scaling $\tau_{\rm exp}\propto t_0/a^2$ is expected with open boundary 
conditions~\cite{Luscher:2011kk}.

%----------------------------------------------------------------------------
\section{Results from lattice simulations}\label{s:res}

On the $\Nf=2$ and $\Nf=0$ ensembles
we measure the following ratios of hadronic scales (cf.~\eq{e:ratio}) 
\begin{eqnarray}
R & = & \sqrt{t_c/t_0}\,,\; \sqrt{t_0}/w_0\,,\; r_0/\sqrt{t_0} \,.
\end{eqnarray}
The scale $t_0$ is defined through~\cite{Luscher:2010iy}
\begin{equation}\label{e:t0}
{\mathcal E}(t_0) = 0.3 \,,\quad {\mathcal E}(t)=t^2\ev{E(x,t)} \,.
\end{equation}
where $E=\frac{1}{4}G_{\mu\nu}^aG_{\mu\nu}^a$ is the action density
of the gauge field smoothed by the Wilson flow 
\cite{Narayanan:2006rf,Luscher:2009eq,Lohmayer:2011si} and
$t$ is the flow time whose mass dimension is $-2$.
Similarly, the scale $t_c$ is defined by the condition
\begin{equation}\label{e:tc}
{\mathcal E}(t_c) = 0.2 \,.
\end{equation}
The scale $w_0$ is defined as~\cite{Borsanyi:2012zs}
\begin{equation}\label{e:w0}
w_0^2{\mathcal E}'(w_0^2) = 0.3 \,,\quad {\cal E}^\prime(t) = \frac{\rmd}{\rmd t}{\cal E}(t) \,.
\end{equation}
The scale $r_0$ is determined
through the condition~\cite{Sommer:1993ce}
\begin{equation}\label{e:r0}
r_0^2F(r_0) = 1.65 \,,
\end{equation}
where the static force $F(r)=V'(r)$ is the derivative of the static
potential $V$.
\begin{figure}[t]\centering
  \includegraphics[width=7cm,clip]{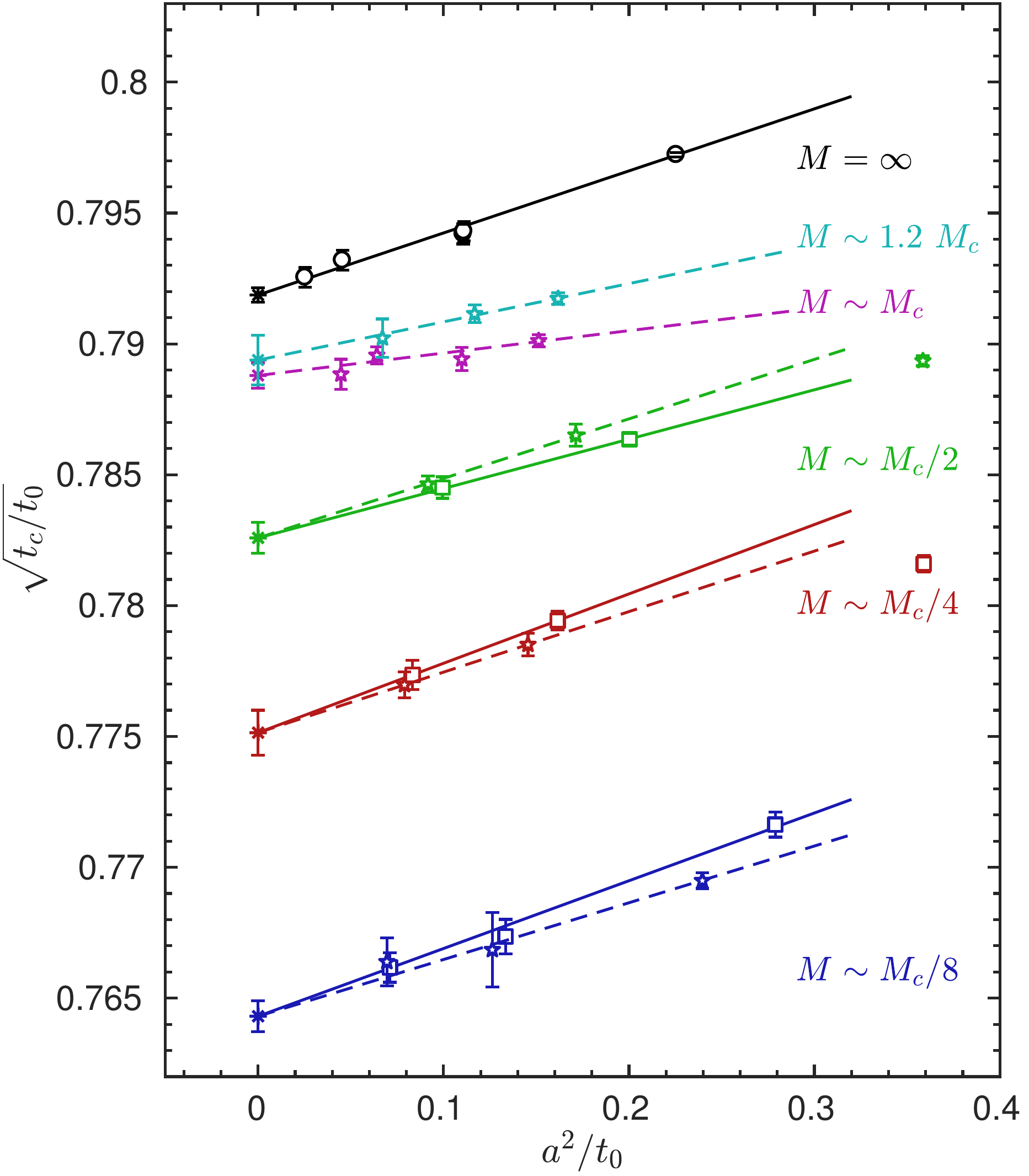}\hfill 
  \includegraphics[width=7cm,clip]{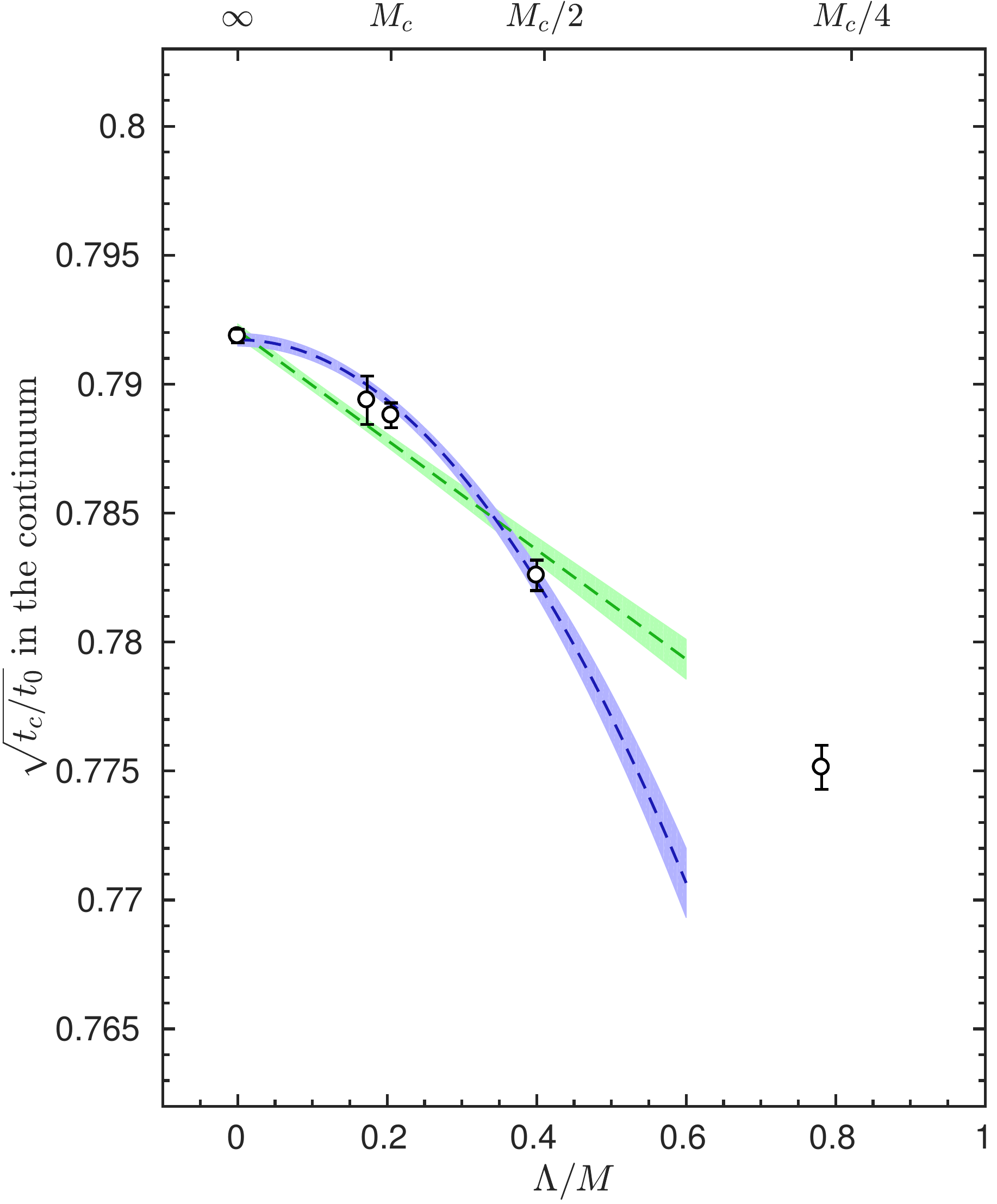}
  \caption{Left: continuum extrapolations of the ratio $R=\sqrt{t_c/t_0}$.
Data are from twisted mass, standard Wilson and pure gauge simulations and are
represented by pentragrams, squares and circles respectively.
The lines show the fits \eq{e:noglobfit}. Dashed lines are for twisted mass 
data and continued lines for Wilson and pure gauge data. 
The continuum extrapolated ratios are shown by the asterisks.
Right: the continuum values plotted against $\Lambda/M$. The dashed line in
the blue band represents the effective theory prediction \eq{e:ratio_EFT}
fitted through points from $M=\infty$ down to $M/\Lambda\ge2.5000$.
The dashed line in the green band is a linear fit in $M^{-1}$.
}
  \label{f:t0tc_noglob}
\end{figure}

For a given value of $M/\Lambda$ and action A (twisted mass, standard Wilson or
for $M=\infty$ pure gauge) we perform continuum extrapolations of the ratios.
From Symanzik's theory we expect O($a^2$) cut-off effects.
We fit our data to
\begin{eqnarray}\label{e:noglobfit}
R(a,M/\Lambda,\rmA) &=& R^{\rm cont}(M/\Lambda) + \frac{a^2}{t_0} c(M/\Lambda, \rmA) \,,
\end{eqnarray}
where the fit parameters are the continuum values $R^{\rm cont}(M/\Lambda)$
and the slopes $c(M/\Lambda, \rmA)$.
In the case where simulations at the same mass were performed with two 
different actions we take a combined continuum limit.
We apply a cut, $a^2/t_0(M)<0.32$, to the data being fitted.
The fits of the ratio $R=\sqrt{t_c/t_0}$ are shown in the left plot of 
\fig{f:t0tc_noglob} and the continuum extrapolated values are listed in
\tab{t:cl}.
In the right plot of \fig{f:t0tc_noglob}
the continuum values $R^{\rm cont}(M/\Lambda)$ are plotted against $\Lambda/M$.
The dashed line in the blue band represents the effective theory prediction
\eq{e:ratio_EFT} fitted through data points from $M=\infty$ down to 
$M/\Lambda\ge2.5000$. It has a good $\chi^2/{\rm dof} = 1.75/2$. 
A linear fit in $M^{-1}$ is shown by the dashed line in the green band and
has a far worse $\chi^2/{\rm dof} = 9.55/2$.
More fits are discussed in Ref.~\cite{Knechtli:2017xgy}. They clearly support
the onset of the effective theory behavior \eq{e:ratio_EFT} 
once data for $M\gtrsim \Mc$ are included in the analysis.

\begin{figure}[t]\centering
  \includegraphics[width=7cm,clip]{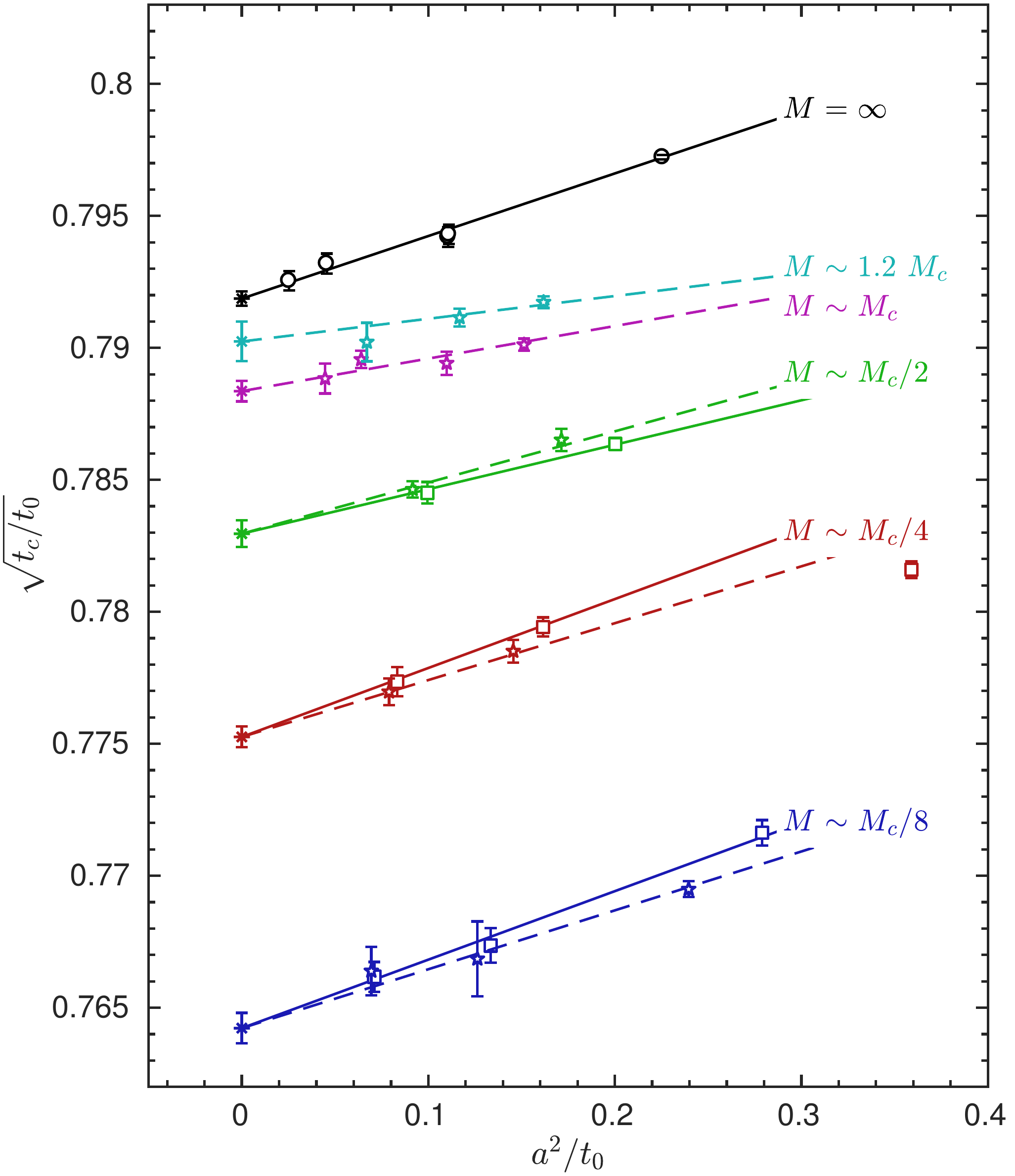}\hfill 
  \includegraphics[width=7cm,clip]{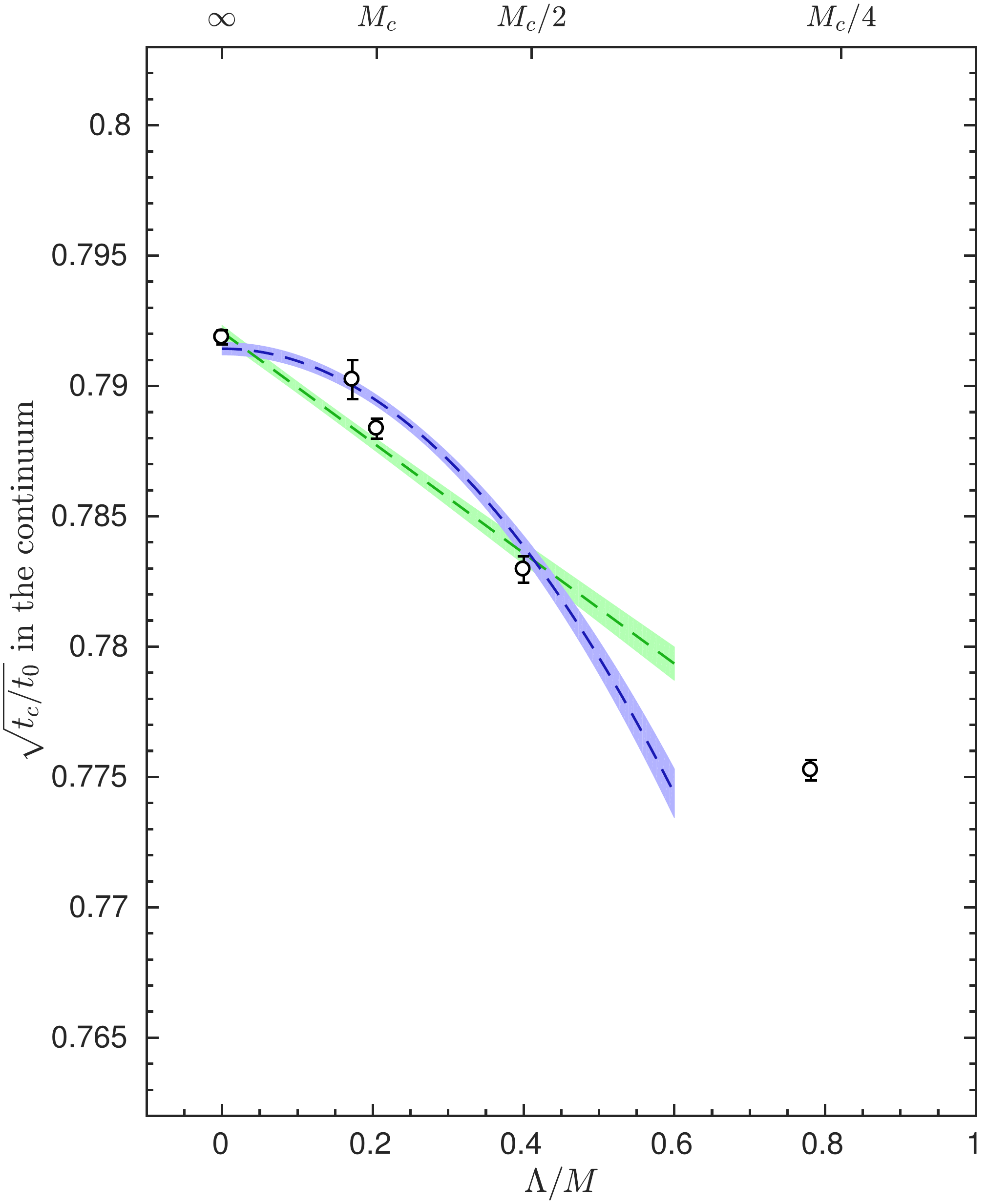}
  \caption{Left: continuum extrapolations of the ratio $R=\sqrt{t_c/t_0}$ using
a global fit \eq{e:globfit}. The symbols and lines are like in 
\fig{f:t0tc_noglob}.
Right: the continuum values plotted against $\Lambda/M$. The lines in the bands
are like in \fig{f:t0tc_noglob}. Note that the continuum data are 
correlated unlike in \fig{f:t0tc_noglob}.}
  \label{f:t0tc_glob}
\end{figure}

For a check we also perform a global fit
\begin{eqnarray}\label{e:globfit}
R(a,M/\Lambda,\rmA) &=&
R^{\rm cont}(M/\Lambda) + \frac{a^2}{8t_0}\left[
c(\rmA) + \alpha(\rmA) \frac{M}{\Lambda} + \beta(\rmA) \frac{M^2}{\Lambda^2}
\right] \,,
\end{eqnarray}
where the slopes are parametrized by the mass-independent coefficient
$c(\rmA)$ and by the parameters $\alpha(\rmA)$, $\beta(\rmA)$ which model the
mass dependence. The coefficient $c(\rmA)$ is the same for twisted mass
and standard Wilson since these two actions are equivalent
for massless quarks. The global fit of the ratio $R=\sqrt{t_c/t_0}$ is shown 
in the left plot of \fig{f:t0tc_glob}. It yields consistent continuum values
with the non-global fit as can be seen from \tab{t:cl}.
In the right plot of \fig{f:t0tc_glob} we show the continuum values
with the quadratic (dashed line in the blue band) and linear (dashed line
in the green band) fits in $M^{-1}$. Note that the errors
of the continuum values are now correlated and this correlation is taken into
account in the fits.
The quadratic fit \eq{e:ratio_EFT} has a correlated 
$\chi^2/{\rm dof} = 16.0/2$ and the linear fit has $\chi^2/{\rm dof} = 7.6/2$.
None of the fits work well although it seems that the quadratic fit
describes better the curvature in the data.
\begin{table}[tph]
\small
 \centering
 \caption{Continuum extrapolated values of the ratios $\sqrt{t_c/t_0}$ and
$\sqrt{t_0}/w_0$. The non-global extrapolations are performed using the fit
\eq{e:noglobfit} and the global extrapolations using the fit \eq{e:globfit}.}
 \label{t:cl}
\begin{tabular}{c c c c c c c}
\toprule
$M/\Lambda$      & $\infty$  & 5.7781     & 4.87      & 2.50      & 1.28        & 0.59 \\
\midrule
\multicolumn{7}{c}{non-global continuum limit} \\
$\sqrt{t_c/t_0}$ & 0.7919(3) & 0.7894(9)  & 0.7888(5) & 0.7826(6) & 0.7751(9)  & 0.7643(6)\\ 
$\sqrt{t_0}/w_0$ & 0.9803(6) & 0.9774(21) & 0.9765(10)& 0.9661(13)& 0.9532(18)  & 0.9311(15)\\
\midrule
\multicolumn{7}{c}{global continuum limit} \\
$\sqrt{t_c/t_0}$ & 0.7919(3) & 0.7902(7)  & 0.7884(4) & 0.7830(5) & 0.7753(4)  & 0.7642(6)\\
$\sqrt{t_0}/w_0$ & 0.9803(6) & 0.9793(17) & 0.9757(9) & 0.9669(11)& 0.9533(9)   & 0.9308(14)\\
\bottomrule
\end{tabular}
\end{table}
\begin{figure}[t]\centering
  \includegraphics[width=7cm,clip]{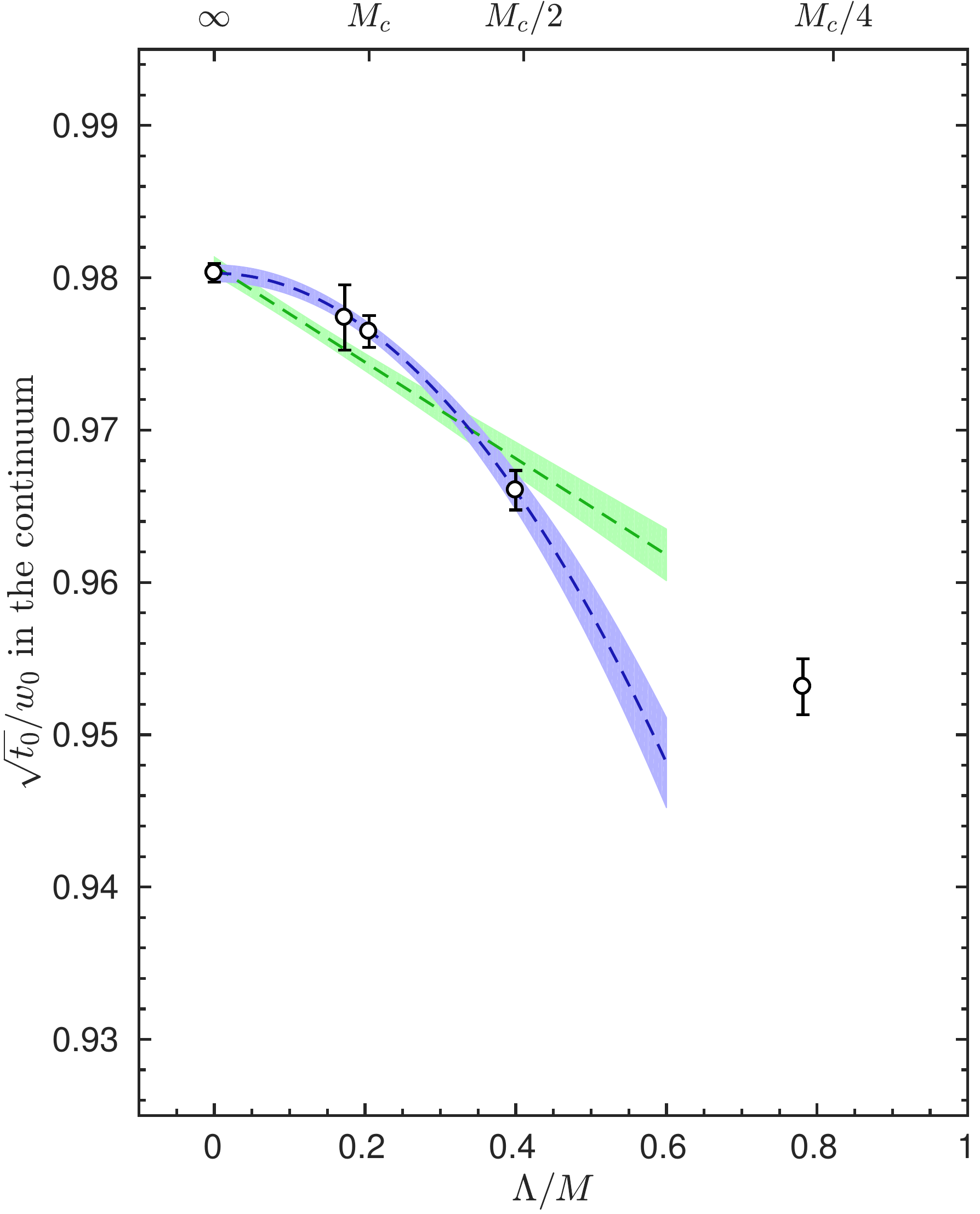}\hfill 
  \includegraphics[width=7cm,clip]{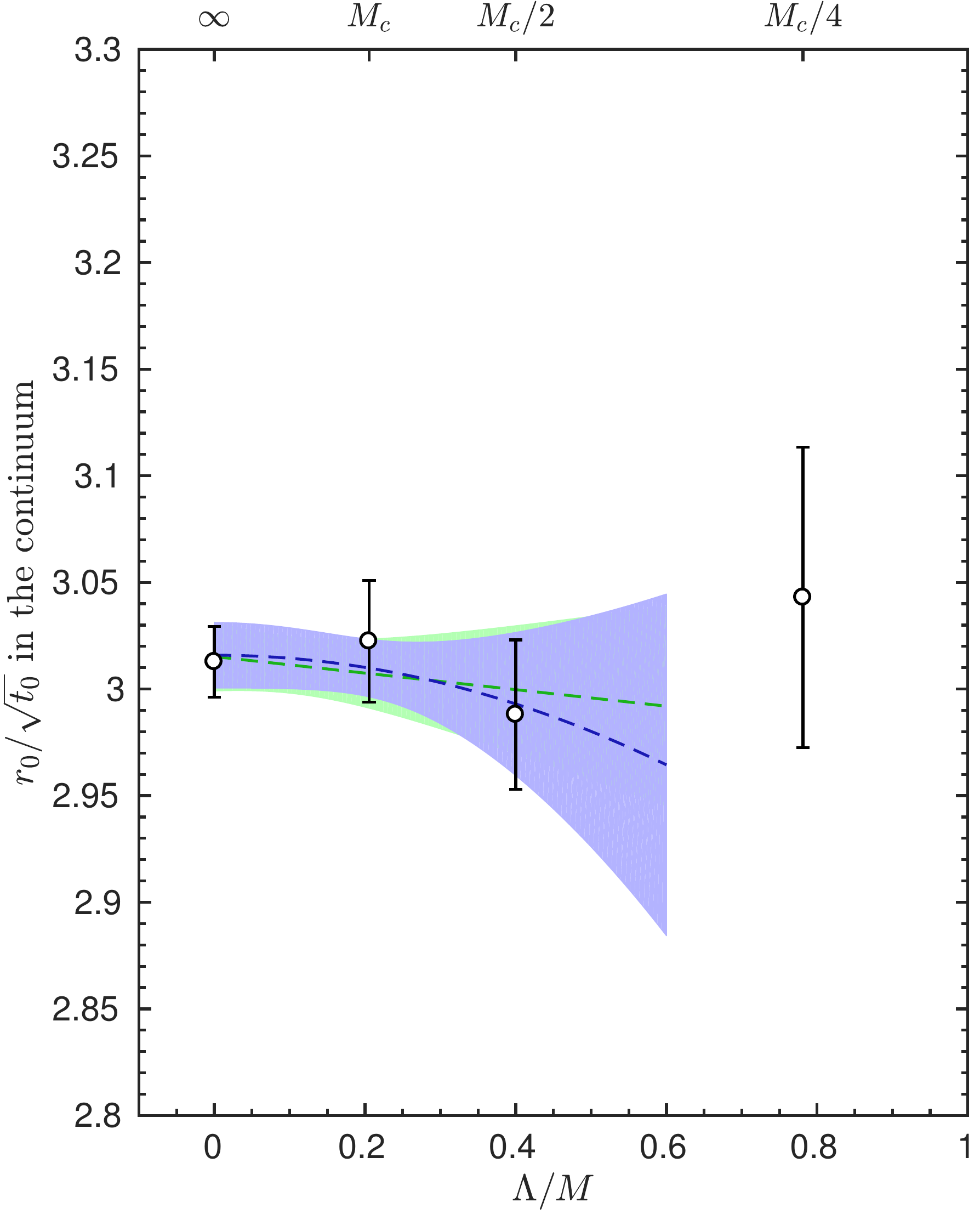}
  \caption{Left: the continuum values of the ratio $R=\sqrt{t_0}/w_0$
determined from the fit \eq{e:noglobfit} 
plotted against $\Lambda/M$. The lines in the bands
are like in the right plot of \fig{f:t0tc_noglob}.
Right: same for the ratio $R=r_0/\sqrt{t_0}$.}
  \label{f:t0w0_r0t0}
\end{figure}

The continuum extrapolated values of the ratios $R=\sqrt{t_0}/w_0$ (left plot) 
and $R=r_0/\sqrt{t_0}$ (right plot) are plotted against $\Lambda/M$ in 
\fig{f:t0w0_r0t0}. They are obtained using the fit \eq{e:noglobfit}.
The ratio $R=\sqrt{t_0}/w_0$ strongly favors the $M^{-2}$ behavior as it is
the case for $R=\sqrt{t_c/t_0}$.
Although a state of the art determination for $r_0$ has been 
used~\cite{Donnellan:2010mx},
the precision of the ratio $R=r_0/\sqrt{t_0}$ is not enough to resolve the 
power corrections.

%----------------------------------------------------------------------------
\section{Conclusions and outlook}\label{s:concl}

We studied the decoupling of a charm quark non-perturbatively in
QCD with two heavy quarks of mass $M$.
By comparing ratios of hadronic flow scales to their values in the
Yang--Mills theory we are able
to measure the effects of a dynamical charm quark which are of 2 permille size.
Our data can be very well fitted by the effective theory prediction for the 
power corrections $\propto M^{-2}$ down to masses $\Mc/2$.

As an outlook, we are computing the effects of a dynamical charm in other 
observables such as
the charm quark mass and charmonium~\cite{Korzec:2016eko} and the
strong coupling from the static force~\cite{Cali:2017brl}.

\noindent
{\bf Acknowledgements.}
The authors gratefully acknowledge the Gauss Centre for Supercomputing (GCS) 
for providing computing time for a GCS Large-Scale Project on the GCS share of 
the supercomputer JUQUEEN at J\"ulich Supercomputing Centre (JSC). GCS is the 
alliance of the three national supercomputing centres HLRS (Universit\"at 
Stuttgart), JSC (Forschungszentrum J\"ulich), and LRZ (Bayerische Akademie der 
Wissenschaften), funded by the German Federal Ministry of Education and 
Research (BMBF) and the German State Ministries for Research of 
Baden-W\"urttemberg (MWK), Bayern (StMWFK) and Nordrhein-Westfalen (MIWF). GM
acknowledges support from the Herchel Smith Fund at the University of Cambridge.
This work is supported by the Deutsche Forschungsgemeinschaft in the SFB/TR 55.

%----------------------------------------------------------------------------

\bibliography{charm}

\begin{thebibliography}{24}

\bibitem{tomaszlat17}
M.~Bruno, M.~Dalla~Brida, P.~Fritzsch, T.~Korzec, A.~Ramos, S.~Schaefer,
  H.~Simma, S.~Sint, R.~Sommer (ALPHA), \emph{{Determination of the Strong
  Coupling Constant by the ALPHA Collaboration}}, in \emph{Proceedings,
  \href{http://inspirehep.net/record/1425631}{35th International Symposium on
  Lattice Field Theory (Lattice2017)}: Granada, Spain}, to appear in EPJ Web
  Conf.

\bibitem{Bruno:2017gxd}
M.~Bruno, M.~Dalla~Brida, P.~Fritzsch, T.~Korzec, A.~Ramos, S.~Schaefer,
  H.~Simma, S.~Sint, R.~Sommer, Phys. Rev. Lett. \textbf{119}, 102001 (2017),
  \texttt{1706.03821}

\bibitem{Bruno:2014ufa}
M.~Bruno, J.~Finkenrath, F.~Knechtli, B.~Leder, R.~Sommer (ALPHA), Phys. Rev.
  Lett. \textbf{114}, 102001 (2015), \texttt{1410.8374}

\bibitem{Knechtli:2017xgy}
F.~Knechtli, T.~Korzec, B.~Leder, G.~Moir (2017), \texttt{1706.04982}

\bibitem{Weinberg:1980wa}
S.~Weinberg, Phys. Lett. \textbf{B91}, 51 (1980)

\bibitem{Cho:1994yu}
P.L. Cho, E.H. Simmons, Phys. Rev. \textbf{D51}, 2360 (1995),
  \texttt{hep-ph/9408206}

\bibitem{Manohar:1997qy}
A.V. Manohar, Phys. Rev. \textbf{D56}, 230 (1997), \texttt{hep-ph/9701294}

\bibitem{Luscher:2010iy}
M.~{L\"uscher}, JHEP \textbf{08}, 071 (2010), [Erratum: JHEP03,092(2014)],
  \texttt{1006.4518}

\bibitem{Sommer:1993ce}
R.~Sommer, Nucl. Phys. \textbf{B411}, 839 (1994), \texttt{hep-lat/9310022}

\bibitem{Frezzotti:2000nk}
R.~Frezzotti, P.A. Grassi, S.~Sint, P.~Weisz (Alpha), JHEP \textbf{08}, 058
  (2001), \texttt{hep-lat/0101001}

\bibitem{Jansen:1998mx}
K.~Jansen, R.~Sommer (ALPHA), Nucl. Phys. \textbf{B530}, 185 (1998), [Erratum:
  Nucl. Phys.B643,517(2002)], \texttt{hep-lat/9803017}

\bibitem{algo:openQCD}
M.~{L\"uscher}, S.~Schaefer, Comput.Phys.Commun. \textbf{184}, 519 (2013),
  \texttt{1206.2809}

\bibitem{Blossier:2012qu}
B.~Blossier, M.~Della~Morte, P.~Fritzsch, N.~Garron, J.~Heitger, H.~Simma,
  R.~Sommer, N.~Tantalo (ALPHA), JHEP \textbf{09}, 132 (2012),
  \texttt{1203.6516}

\bibitem{Fritzsch:2012wq}
P.~Fritzsch, F.~Knechtli, B.~Leder, M.~Marinkovic, S.~Schaefer, R.~Sommer,
  F.~Virotta, Nucl. Phys. \textbf{B865}, 397 (2012), \texttt{1205.5380}

\bibitem{Wolff:2003sm}
U.~Wolff (ALPHA collaboration), Comput.Phys.Commun. \textbf{156}, 143 (2004),
  \texttt{hep-lat/0306017}

\bibitem{Schaefer:2010hu}
S.~Schaefer, R.~Sommer, F.~Virotta (ALPHA Collaboration), Nucl.Phys.
  \textbf{B845}, 93 (2011), \texttt{1009.5228}

\bibitem{Luscher:2011kk}
M.~L{\"u}scher, S.~Schaefer, JHEP \textbf{1107}, 036 (2011), \texttt{1105.4749}

\bibitem{Narayanan:2006rf}
R.~Narayanan, H.~Neuberger, JHEP \textbf{03}, 064 (2006),
  \texttt{hep-th/0601210}

\bibitem{Luscher:2009eq}
M.~L{\"u}scher, Commun. Math. Phys. \textbf{293}, 899 (2010),
  \texttt{0907.5491}

\bibitem{Lohmayer:2011si}
R.~Lohmayer, H.~Neuberger, PoS \textbf{LATTICE2011}, 249 (2011),
  \texttt{1110.3522}

\bibitem{Borsanyi:2012zs}
S.~Borsanyi, S.~D{\"u}rr, Z.~Fodor, C.~Hoelbling, S.D. Katz, S.~Krieg,
  T.~Kurth, L.~Lellouch, T.~Lippert, C.~McNeile et~al., JHEP \textbf{09}, 010
  (2012), \texttt{1203.4469}

\bibitem{Donnellan:2010mx}
M.~Donnellan, F.~Knechtli, B.~Leder, R.~Sommer, Nucl.Phys. \textbf{B849}, 45
  (2011), \texttt{1012.3037}

\bibitem{Korzec:2016eko}
T.~Korzec, F.~Knechtli, S.~Cali, B.~Leder, G.~Moir, \emph{{Impact of dynamical
  charm quarks}}, in \emph{{Proceedings, 34th International Symposium on
  Lattice Field Theory (Lattice 2016): Southampton, UK, July 24-30, 2016}}
  (2016), \texttt{1612.07634}

\bibitem{Cali:2017brl}
S.~Cal{\`i}, F.~Knechtli, T.~Korzec, H.~Panagopoulos, \emph{{Charm quark
  effects on the strong coupling extracted from the static force}} (2017), in
  \emph{Proceedings, \href{http://inspirehep.net/record/1425631}{35th
  International Symposium on Lattice Field Theory (Lattice2017)}: Granada,
  Spain}, to appear in EPJ Web Conf., \texttt{1710.06221}

\end{thebibliography}

%%%%%%%%%%%%%%%%%%%%%%%%%%%%%%%%%%%%%%%%%%%%%%%%%%%%%%%%%%%%%%%%%%%%%%%%%%%%%
\end{document}